\documentclass[aps,prl,reprint,noshowpacs,superscriptaddress,floatfix,letterpaper,notitlepage]{revtex4-1}
\usepackage{amsmath,amssymb,amsbsy,amsfonts,amsthm,bbm,bm,mathtools,mathrsfs}
\usepackage{color}
\usepackage{physics}
\usepackage[dvipsnames]{xcolor}
\pagecolor{white}
\definecolor{LapisLazuli}{RGB}{47, 102, 169}
\usepackage[colorlinks=true,citecolor=LapisLazuli,linkcolor=LapisLazuli,urlcolor=LapisLazuli]{hyperref}
\newcommand{\jrg}[1]{\textcolor{black}{#1}}

\begin{document}

\title{Stochastic paths controlling speed and dissipation}

\author{Rebecca~A.~Bone}
\affiliation{Department of Chemistry,\
  University of Massachusetts Boston,\
  Boston, MA 02125, USA
}

\author{Daniel~J.~Sharpe}
\author{David~J.~Wales}
\affiliation{Department of Chemistry,\
University of Cambridge,\
Lensfield Road, Cambridge CB2 1EW, UK
}

\author{Jason~R.~Green}
\email[]{jason.green@umb.edu}
\affiliation{Department of Chemistry,\
  University of Massachusetts Boston,\
  Boston, MA 02125, USA
}
\affiliation{Department of Physics,\
  University of Massachusetts Boston,\
  Boston, MA 02125, USA
}

\begin{abstract}

Near equilibrium, thermodynamic intuition suggests that fast, irreversible processes will dissipate more energy and entropy than slow, quasistatic processes connecting the same initial and final states.
\jrg{Here, we test the hypothesis that this relationship between speed and dissipation holds for stochastic processes far from equilibrium.}
To analyze \jrg{these processes on finite timescales}, we derive an exact expression for the path probabilities of continuous-time Markov chains from the path summation solution of the master equation.
Applying this formula to a model for nonequilibrium self-assembly, we show that \jrg{more speed can lead to less dissipation} 
when there are strong nonequilibrium currents.
\jrg{In the model, the relative energies of the initial and target states control the speed, and the nonequilibrium currents of a cycle situated between these states control the path-level dissipation.}
\jrg{This model serves as a minimal prototype for designing kinetics to sculpt the nonequilibrium path space, so that faster structure-forming paths dissipate less.}

\end{abstract}

\maketitle

Biological systems have evolved the ability to balance structure formation on a
particular time scale with the thermodynamic costs of dissipated heat, entropy production, and wasted free energy~\cite{Marsland2017,Mishra2021}.
Prominent examples include nonequilibrium self-assembly and protein folding, where structure forms as the system flows down an energy gradient.
\jrg{The usefulness of these processes often requires the structure be formed on a particular timescale~\cite{vanRossum2017, Bishop2009}, with the energy efficiency being a consequence of the requisite speed~\cite{Bryant2020, Song2020,Nicholson2020}.}
\jrg{Equilibrium thermodynamics suggests that faster physical processes will dissipate more than a slow, reversible process~\cite{callen85,Falasco2020}.}
\jrg{However, if this relationship} holds at smaller length and time scales, systems should dissipate more to actuate structure formation on a specified timescale~\cite{Grzybowski2009, Whitelam2009, Murugan2012, Fang2019}.
A complication is that structure formation often proceeds through a set of intermediate, misassembled, or misfolded states that separate the initial and target states~\cite{Komine2019, vanRavensteijn2020}.
\jrg{When open to external sources of energy and matter, kinetic trapping and dissipative cycles~\cite{Rao2018} are prevalent}~\cite{Whitelam2015, Zhou2021, Penocchio2019}, making it an open question how physical systems use time and energy in the formation of a target structure.

\jrg{Dissipative cycles in open chemical reaction networks are being designed that control the timing of material structure formation~\cite{Boekhoven2015,Zhou2021}.}
\jrg{These chemically active materials are an example of a broader idea to leverage the richer set of structures that can be formed outside of chemical, thermal, and diffusive equilibria~\cite{Whitelam2014, Yasui2015, Tagliazucchi9751, Nguyen2016}.}
The structural and mechanical properties of the resulting materials can depend on the order of the protocol used in preparation~\cite{Macdougall2017, Diggle2021},
\jrg{which motivates a statistical treatment of the stochastic paths} to accurately represent the process~\cite{Nicholson2016learning, Nicholson2018effects, Sharpe2019, Swinburne2020}.
Stochastic paths of assembly histories provide quantitative dynamical information~\cite{GreenCGS13, Nicholson2019} and can vary significantly in the associated dissipation, kinetics, and resulting structures~\cite{Altaner2012}\jrg{, raising the question of the relationship between speed and dissipation}.
Markov models, parameterized based on deterministic trajectories~\cite{Chodera2007,Prinz2011} or kinetic theory~\cite{wales2003energy}, are a popular approach to model stochastic processes.
Stochastic paths are numerically sampled or explicitly enumerated~\cite{Trygubenko2006, Nicholson2019, Wang1996, Feng2010, Nicholson2018}.
\jrg{However, to analyze the stochastic thermodynamics of transient assembly processes, we need a means of computing dissipation with an explicit accounting of time.}

In this Letter, we derive an explicit form of the path probability of any continuous-time Markov process. \jrg{This closed form expression provides a direct method for computing the average speed and dissipation of nonequilibrium paths over a fixed period of time.}
Specifically, we derive the probability that a time-ordered sequence of states (a path, $\mathcal{C}_{n}=x_{0},x_{1},\ldots,x_{n}$) traversed by an ensemble of stochastic trajectories (a time-ordered sequence of states and stochastic transition times, $\mathcal{T}_{n}=x_{0},t_{0};x_{1},t_{1};\ldots ;x_{n},t_{n}$) in a fixed observation time $\tau\geq t_n$~\cite{Sun2006, Bai2015JPCL, Bai2015JPCA, Bai2017}.
\jrg{The fixed time for the process constrains the possible paths included in a path summation~\cite{Helbing1, Helbing2} and, in turn, the thermodynamic cost.
Leveraging this approach, we quantitatively analyze whether ``more haste'' necessarily means ``more waste'' for potentially transient stochastic paths.}

\textit{Contracted path probability.--} 
We consider Markov jump processes~\cite{Gillespie1991} for mesoscopic systems with a discrete set of $N$ states.
These stochastic dynamics have been used to describe self-assembly~\cite{D'Orsogna2012}, but also quantum dots~\cite{Head-Marsden2021} and molecular motors~\cite{Seiferth2020}.
The physical dynamics are represented by a collection of transition rates, $w(y|x)$ for a jump from state $x$ to state $y$ and the total escape rate $w_{x} = \sum_{y\neq x}w(y|x)$ from $x$.
The dynamics of occupation probabilities for each state are described by the master equation~\cite{Gillespie1991}.
Its path summation solution~\cite{Weber2017} gives the marginal probability of state \jrg{$x_{i}$,
\begin{align}
  p(x_i,\tau) & = \sum_{n=0}^\infty\sum_{\mathcal{C}_n}\mu(\mathcal{C}_n=x_0,x_1,\ldots,x_i,\tau),
\end{align}}
in terms of the joint probability $\mu$ that the system takes \jrg{paths $\mathcal{C}_{n}$ of any number of jumps $n$ that end in $x_{i}$ after the time $\tau=t-t_{0}$.
Because of the stochastic transition times, $\mathcal{T}_n$, some paths remain in the final state for $t-t_{n}\geq 0$, but others may not reach $x_i$ or may transition out of $x_i$ on the chosen time $\tau$.
As a consequence, fixing the time alters nonequilibrium ensemble averages over paths.}

Our main technical result is an exact expression for this contracted path
probability, $\mu(\mathcal{C}_{n},\tau)$ (Supplementary Materials, SM). 
Difficulties in deriving this expression have been discussed previously~\cite{Sun2006, FlomenbomKS2006, SunReply2006}.
The probability, $\mu(\mathcal{C}_{n},\tau)=p(\tau \vert
\mathcal{C}_{n})p(\mathcal{C}_{n})$, is expressed in terms of the probability of a path $p(\mathcal{C}_{n}) = p(x_{0}) \prod_{i=1}^{n} w(x_{i} \vert x_{i-1})/w_{x_{i-1}}$ and its stochastic time sequence,
\begin{equation}
\label{eq:contrpath}
p(\tau \vert \mathcal{C}_{n}) =
\prod\limits_{i=0}\limits^{n-1} w_{x_{i}} \sum\limits_{j=1}\limits^{n'} \nu_{j} f_{j}^{(0)}
\sum\limits_{l=1}\limits^{m_{j}} \binom{m_{j}-1}{l-1} \left( - \tau \right)^{m_{j}-l} d_{j}^{(l-1)},
\end{equation}
which consists of three functions.
The first, $\nu_{j}$, 
accounts for the multiplicity $m_j$ of the $j$th out of $n'$ unique escape rates along a path.
The second is a function proportional to the exponential distribution of waiting times in each state, $f_{j}^{(0)} \propto e^{-w_{x_{j}}\tau}$.
The third function, $d_{j}^{(l-1)}$ (SM), is a nontrivial combinatoric function of the escape rate multiplicities that derives from the nested convolutions of the waiting time distributions; these combinatorics were previously pointed out as a particular challenge in deriving this closed-form solution~\cite{FlomenbomKS2006}.

This expression for the path probability has advantages for the stochastic thermodynamics of processes constrained by time.
\jrg{It allows direct quantification of the relative importance of paths with respect to their stochastic time sequences, making it useful for extracting insights into dynamical mechanisms. For example, the magnitude of the probability can be used to assess the relative importance of competing mechanisms.}
This formula simplifies to known expressions~\cite{Helbing1} when the escape rates are distinct,
\begin{align}
\mu(\mathcal{C}_{n},\tau) =
p(x_{0},t_{0}) &\prod_{i=1}^{n} w(x_{i} \vert x_{i-1})\\\nonumber
  &\sum_{j=0}^{n} \frac{e^{-w_{x_{j}}\tau}}{\prod_{k=1,\neq j}^{n} (w_{x_{k}}-w_{x_{j}})^{m_{k}}},
\end{align}
and when the escape rates $w$ along a path are identical, SM.
In the former case, the sum is over the $f_{j}^{(0)}$ for each escape rate $j$ along the path (SM).

\begin{figure}[!t]
\centering
\includegraphics[width=1\columnwidth]{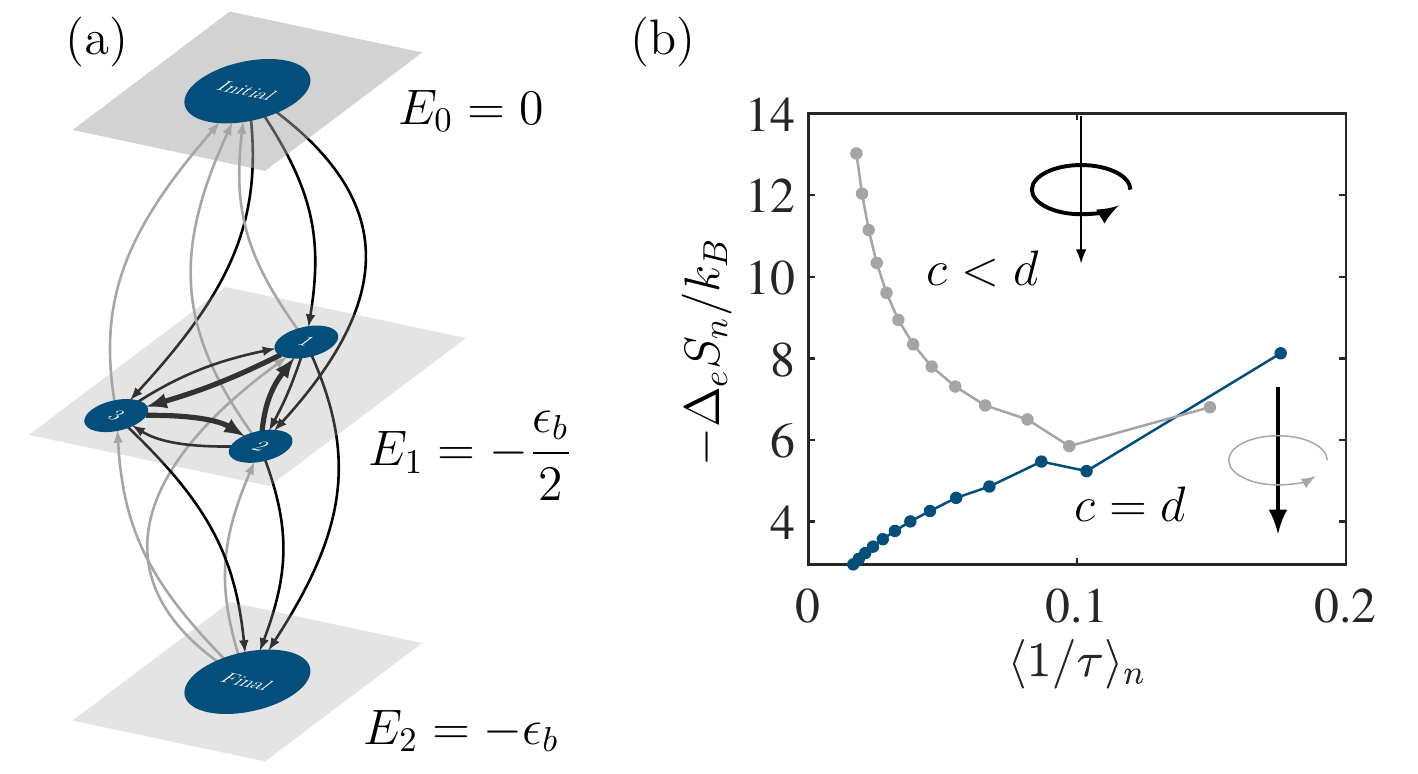}

\caption{\label{fig1}
\textit{Completing a nonequilibrium process more quickly can result in more or less dissipated entropy on average.}
(a) Model with three energy levels in which a potentially dissipative cycle is intermediate between initial (high energy) and final (low energy) states.
\jrg{Transition rates up the energy gradient (gray) are a function $\exp(-\beta\epsilon_{b}/2)$ of the effective binding energy $\beta\epsilon_{b}$ in units of $k_{B}T$.
Those down the energy gradient (thin black) represent the concentration of monomers, $c$.}
Transitions around the cycle can dominate the entropy flow. 
Clockwise (thick black) transitions with rate constant $d$ can be tuned relative to clockwise (thin black) with rate constant $c$.
(b) Dissipation $-\Delta_e S_n/k_{B}$ versus speed $\langle\tau^{-1}\rangle_n$, both conditioned on path length $n$, for the model in (a). 
\jrg{When $c=d$ (blue), dissipation increases with the average speed of paths that connect unassembled and assembled states. 
However, when $c\leq d$ (gray), dissipation decreases with the
average rate of path completion.}
}

\end{figure}

\begin{figure}[t!]

\centering
\includegraphics[width=1\columnwidth]{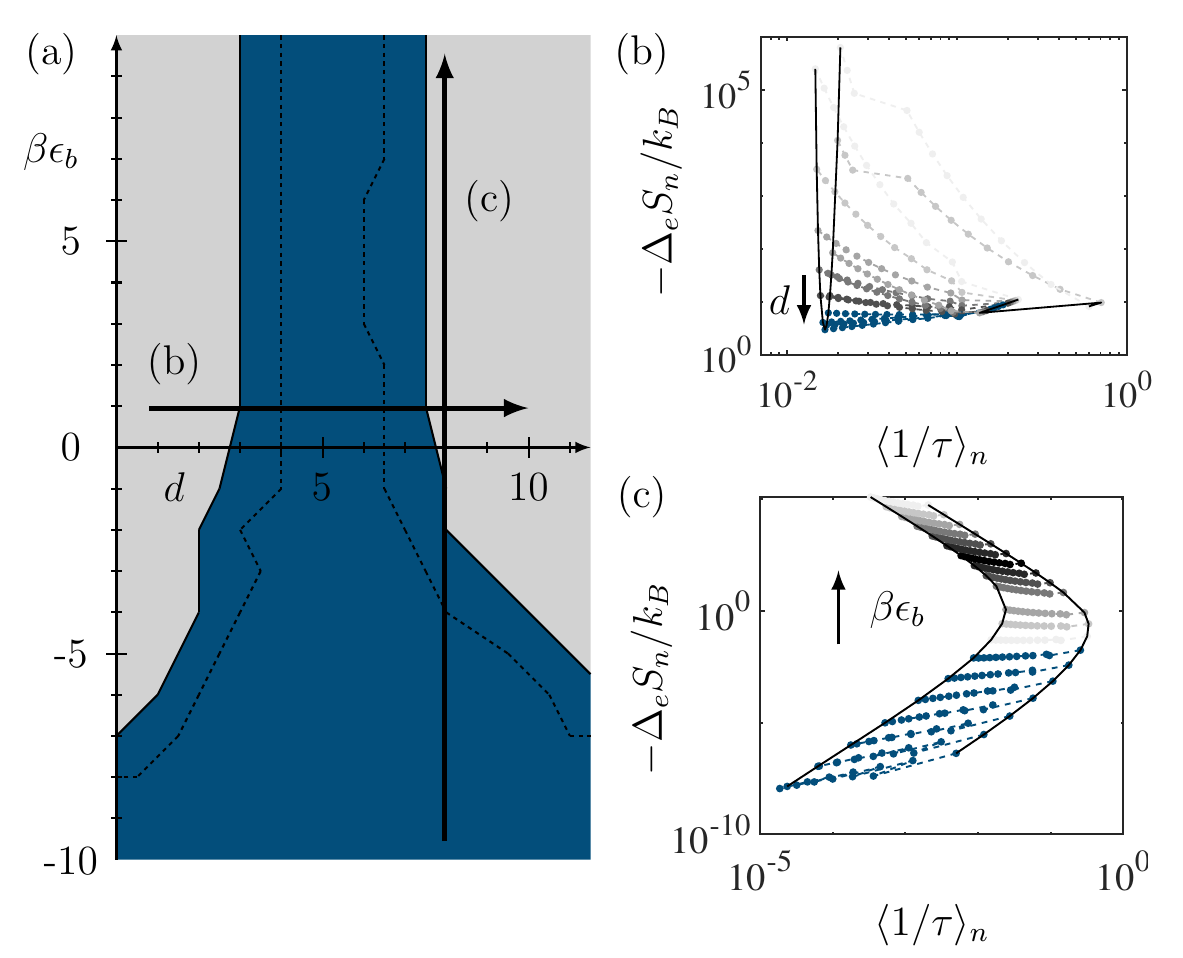}

\caption{\label{fig2}
\textit{Model parameters control whether faster paths dissipate more or less entropy on average.}
(a) Diagram mapping the sign of the slope from $-\Delta_{e} S_{n}/k_{B}$ as a function of
$\langle 1 / \tau \rangle_{n}$ for a cross section of the parameter space
$(\beta\epsilon_b, d \mid c=5)$. 
For $\beta\epsilon_b > 0$, the dissipation
increases with speed when $c\approx d$ (blue). 
For $\beta\epsilon_b < 0$, the
width of this region grows with increasingly negative $\beta\epsilon_b$.
Arrows indicate the parameter sweeps shown in (b-c). 
(b) When paths are resolved by the number of jumps between the initial and final states \jrg{(and averaged)},
the entropy dissipated to the surroundings, $-\Delta_{e} S_{n}/k_{B}$, can increase or decrease as a function of the mean rate of path completion $\langle 1 / \tau \rangle_{n}$. 
\jrg{The slope is controlled by the current around the cycle, which we tune the cycle by sweeping $d\in [0.5,10]$ in increments of 0.5 at fixed $c=5$ and $\beta\epsilon_{b}=1$.}
Dissipation increases with the rate of path completion
(blue) for $c\approx d$ but decreases sharply when $|c-d|\gg 0$ (grayscale). 
Solid lines connecting the slowest paths for the parameter sweep show a parabolic trend (black). 
(c) Dissipation, $-\Delta_{e} S_{n}/k_{B}$, as a function of the mean rate of path completion $\langle 1 / \tau \rangle_{n}$ resolved \jrg{(averaged)} by the path length $n$. 
The slope is controlled by sweeping $\beta\epsilon_{b}\in[-10,10]$ in
increments of 1 with fixed $c=5$ and $d=8$.}

\end{figure}

\jrg{With the general expression for the path probability, $\mu(\mathcal{C}_{n},\tau)$, we consider two ensemble average observables, one for speed and one for dissipation, for the paths from an initial state $x_0$ to a target state $x_i$.}
The amount of time it takes a stochastic trajectory on average to follow a path
$\mathcal{C}_{n}$ is the path occurrence time $\langle \tau
\rangle_{\mathcal{C}_{n}}=\sum_{i=0}^{n} w_{x_{i}}^{-1}$, the cumulative mean
of the independent and exponentially distributed waiting times along the path.
Its inverse $1/\langle \tau \rangle_{\mathcal{C}_{n}}$ is a measure of the ``speed'' at which the system traverses \jrg{a single path}.
\jrg{We first analyze paths of length $n$ connecting two states, measuring the speed with the ensemble average for a given $n$: $\langle 1 / \tau \rangle_{n} := \sum_{\mathcal{C}_n} \langle \tau \rangle_{\mathcal{C}_{n}}^{-1} \mu(\mathcal{C}_{n},\langle \tau \rangle_{\mathcal{C}_{n}})$, and evaluating the path probability at the path occurrence time, $\langle \tau \rangle_{\mathcal{C}_{n}}$.}

Stochastic thermodynamics has well-established measures of entropy dissipation at the path level~\cite{Vandenbroeck2015}.
Assuming local detailed balance~\cite{Maes2021}, the entropy change of the equilibrium reservoirs mediating the fluctuating dynamics puts a constraint on the asymmetry of the transition rates: \jrg{$-s_{e}[\mathcal{C}_{n}]/k_{B}=\sum_{i=0}^{n-1} \ln w(x_{i+1} \vert x_{i})/w(x_{i} \vert x_{i+1})$}.
\jrg{This entropy flow is interpreted as the amount of entropy dissipated from the system to the surroundings in traversing a path~\cite{seifert2012stochastic, Falasco2021}.}
When the transition rates are exponentially related to the energy, $-s_{e}[\mathcal{C}_{n}]/k_{B}$ is the energy exchanged as heat between the system and surroundings, $k_{B}\ln w(x,y)/w(y,x) = q(x,y)$~\cite{Horowitz2015}.
\jrg{For the set of paths of a given $n$, we take the ``dissipation'' to be the entropy dissipated averaged over paths of length $n$, $-\Delta_{e} S_{n}/k_{B}=-\sum_{\mathcal{C}_{n}} (s_{e}[\mathcal{C}_{n}]/k_{B})\mu(\mathcal{C}_{n},\langle \tau \rangle_{\mathcal{C}_{n}})$.}

\textit{Model system.--} \jrg{Equipped with our contracted path probability, we built a minimal model to test the hypothesis that faster paths will dissipate more, Fig.~\ref{fig1}(a).}
The model is an adaptation of Onsager's three-state cycle, which he used for illustrating detailed balance (breaking)~\cite{Onsager1931Pt1}.
\jrg{It is a discrete state Markov model in which paths that lead downhill when $\epsilon_{b}>0$ (or uphill if $\epsilon_{b}<0$) in energy connect an initial state to a final state, with an impeding dissipative cycle.}
Computationally, we evaluate the path ensemble averages of speed and dissipation by enumerating paths up to $n=15$ that connect the monomer and assembled states.
Taking the initial marginal probability $p(x_{0}=1)=1$, we compute the probability of each with our exact expression, but we could also compute the probability of preferred paths (e.g., high probability paths accounting for specified percentage of probability flux) for larger networks using a depth-first search procedure with path-culling criteria~\cite{Helbing2}.

\jrg{The model 
has two parameters representing common experimental control variables: the relative stability of the monomers and assembled structure (the binding energy, $\epsilon_b$) and the concentration of the monomeric units, $c$.}
\jrg{The concentration of monomers, $c$, and the control parameter for dissipation, $d$, determine the preferred jump direction around the cycle. They also control the dissipation of traversing a single step of the cycle $\pm \ln c/d$, which is zero when $c=d$.}

By design, these parameters give control over both the speed of paths connecting the high (low) energy initial state and the low (high) energy final state and the associated dissipation.
\jrg{With increasing path length, $n$, the entropy dissipated $-\Delta_{e} S_{n}/k_{B}$ can increase or decrease, depending on the exact values of the parameters, SM~Fig.~3.
However, the speed $\langle 1 / \tau \rangle_{n}$ decreases with $n$, SM~Fig.~3.
(We note, however, that this condition may not hold in networks whose escape rates are not similar in magnitude.)
Varying these control parameters also modulates the competition between the speed of path completion and the associated entropy dissipated to show that faster paths do not necessarily dissipate more.
Fig.~\ref{fig1}(b) shows that in kinetic networks of this type, the dissipation $-\Delta_{e} S_{n}/k_{B}$ can be an increasing function of the speed $\langle 1/\tau\rangle_n$ -- as one would expect from equilibrium thermodynamics -- when there is no preferred direction among the intermediate states.
However, when there are nonequilibrium currents and a strongly preferred direction among the intermediate states, then faster paths dissipate less.}

To understand the mechanism of this nonequilibrium behavior, consider when the concentration parameters are nearly equal, $c\approx d$.
\jrg{In this case, the energy gradient determines the dependence of dissipation on the timescale of path completion, SM~Fig.~4(c-e)}.
\jrg{Only the sequence of states down the energy gradient
contribute to the path entropy flow $-s_{e}[\mathcal{C}_{2}]/k_{B}=2\ln (c)+\beta\epsilon_{b}$.}
\jrg{(Similar behavior is found when cycle transitions are removed, SM~Fig.~4(c-e).)}
\jrg{Longer paths, which necessarily include additional jumps around the cycle, take longer but do not dissipate more.
Consequently, the ensemble averages $-\Delta_{e} S_{n}/k_{B}$ and $\langle 1/\tau\rangle_n$ are positively correlated because they both decrease with path length $n$.}
Now, when $c\neq d$ and paths are long enough that jumps on the cycle are significant, each transition adds $\pm \ln c/d$ to the entropy flow, SM~Fig.~2(b).
(Contributions to the entropy flow in one direction of the cycle can be negated by subsequent transitions in the reverse direction, so only net transitions on the cycle contribute.)
Paths that make $K$ transitions around the cycle in a particular direction will dissipate an additional $\pm K\ln c/d$.
\jrg{The average entropy dissipated increases with path length $n$.
As a result, it is negatively correlated with the mean path occurrence time $\langle \tau \rangle_{n}$, which increases linearly with $n$, SM~Fig.~2(a).}

\jrg{Because of this balance between dissipation (traversal of the cycle) and speed (traversal of the energy gradient), there are regions of parameter space where the dissipation is an increasing or decreasing function of the rate of path completion, Fig.~\ref{fig2}.}
\jrg{Transitions between these regimes -- where equilibrium thermodynamic intuition does and does not hold -- are controlled by the relative magnitudes of the cycle transition rates, $c/d$, and the relative binding energy, $\beta\epsilon_{b}$.}
\jrg{Dissipation and speed are positively correlated when $c\approx d$, when jumps around the cycle do not significantly increase dissipation.}
Loosely, increasing the speed only increases the dissipation when the cycle is weakly dissipative, $c\approx d$.

\begin{figure}[t!]

\centering
\includegraphics[width=1.0\columnwidth]{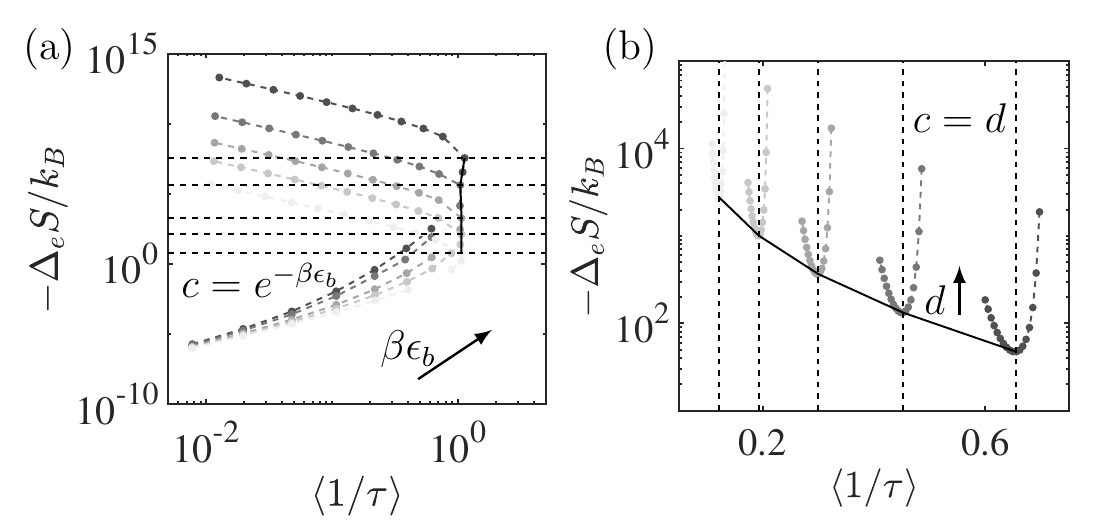}

\caption{\label{fig3} 
\textit{Speed and dissipation extrema in the parameter space.}
(a) \jrg{Entropy dissipated $-\Delta_{e} S/k_{B}$ versus speed $\langle 1 / \tau \rangle$ averaged over all paths varying} $\beta\epsilon_{b}\in [-10,10]$ with fixed $c=5$ and $d=\lbrace 0.5,0.6,1,1.5,4\rbrace$. Darker gray indicates larger values of $d$. Dashed horizontal lines indicate $c=e^{-\beta\epsilon_{b}}$ for each $d$. Solid black line connects vertices for different $d$ values.
(b) Ensemble level dissipation $-\Delta_{e} S/k_{B}$ versus speed $\langle 1 / \tau \rangle$ for $d\in[1.5,10]$ in increments of 0.5 and other parameters fixed at $c=5$ and $\beta\epsilon_{b}\in[1,4]$ in increments of 1 (darker gray indicates lower value of $\beta\epsilon_{b}$). Dashed vertical lines indicate $c=d$ for each fixed $\beta\epsilon_{b}$ value. Solid black line connects vertices for different $\beta\epsilon_{b}$ values.
}

\end{figure}

\jrg{Further, the thermodynamic stability of the assembled state (determined by the binding energy $\beta\epsilon_{b}$ has an effect on where exactly this border between dissipative and weakly/non dissipative cycles lies in parameter space.}
Widening the energy-level gap when $\beta\epsilon_{b}<0$ increases the width of the area in $\vert \log c/d \vert$ parameter space in which this intuition is correct, Fig.~\ref{fig2}(a)(blue).
Under these conditions, paths are uphill in energy ($\beta\epsilon_{b}<0$), so an increasing amount of energy is required to force the system into the energetically unfavorable assembled state.
These paths are then generally lower in probability with less entropy dissipation than their counterparts with higher $\beta\epsilon_{b}$, SM~Fig.~3(d).
\jrg{Further, because the probability decays (exponentially with $n$) more quickly than the entropy dissipation increases (linearly) with $n$,} there is a decrease in the average dissipation with $n$ and a positive correlation between speed and dissipation for sufficiently negative $\beta\epsilon_{b}$, Fig.~\ref{fig2}(b).
\jrg{This result suggests that there is a limit to the size of the energy gap that can exist between states of a dissipative self-assembly system for a given $c$ and $d$.}
\jrg{Beyond this limit, the system requires energy of a similar magnitude (rather than dissipating) in order to drive it into the energetically unfavorable assembled state~\cite{Tagliazucchi9751}.
The same behavior is observed when the cycle transitions are removed and the binding energy is negative, SM~Fig.~4(c-e).}

\begin{figure}[t]

\centering
\includegraphics[width=1\columnwidth]{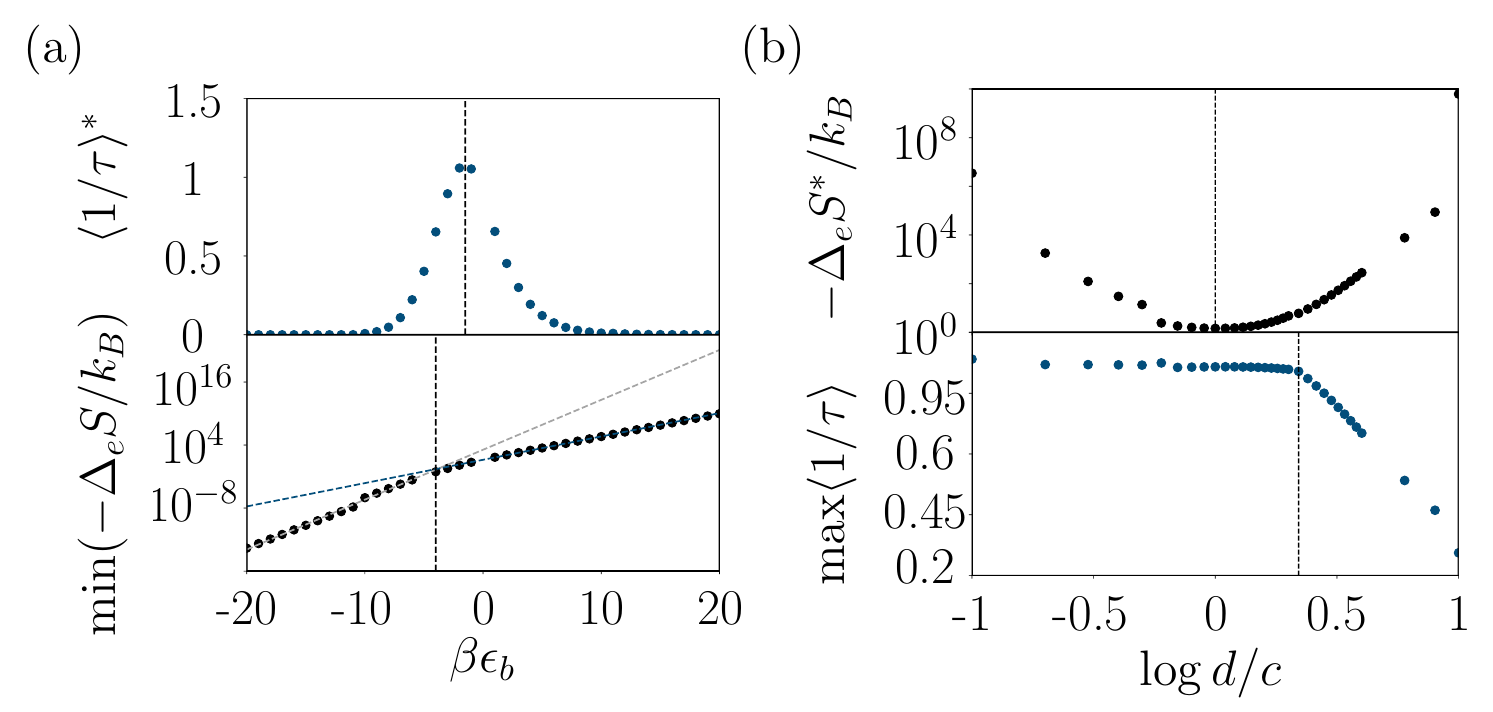}

\caption{\label{fig4}
\textit{Maximizing speed (minimizing dissipation) in the parameter space does not simultaneously minimize dissipation (maximize speed).}
\jrg{(a) Minimum entropy dissipated on average (bottom) and associated speed $\langle 1/\tau\rangle^* := \langle 1/\tau\rangle|_{\min(-\Delta_e S/k_B)}$ (top) for $\beta\epsilon_{b}\in [-10,10]$ 
for fixed $c=5$.}
Dashed lines are exponential trends for $\beta\epsilon_{b} > -\ln c$ (blue) and $\beta\epsilon_{b} < -\ln c$ (gray).
(b) \jrg{Maximum average speed (bottom) and associated average entropy dissipated $-\Delta_e S^*/k_B := -\Delta_e S/k_B|_{\max\langle 1/\tau\rangle}$ (top) for $d\in [0.5,10]$ 
and fixed $c=5$.}
}

\end{figure}

These observations for paths conditioned on their length translate into the average speed, $\langle 1 / \tau \rangle$, and dissipation, $-\Delta_{e} S/k_{B}$, over the entire ensemble of paths, Fig.~\ref{fig3}.
So, we systematically varied the model parameters and identified conditions where speed is maximal and dissipation is minimal.
Scanning values of $\beta\epsilon_{b}$ with fixed $c$ and $d$, there is a maximum average speed located at $c=e^{-\beta\epsilon_{b}}$, when jumps up and down the energy gradient have the same transition rate, Fig.~\ref{fig3}(a).
Scanning $d$ with fixed values of $\beta\epsilon_{b}$ and $c$, there is a minimum average entropy dissipation located at $c=d$, Fig.~\ref{fig3}(b).
These extrema are also apparent in Fig.~\ref{fig2}(b,c).
\jrg{Their locations mark the transition of ensemble averages between regimes of positive and negative correlation between speed and dissipation.}

Extremizing with respect to the model parameters, we find that minimizing dissipation does not simultaneously maximize speed or vice-versa.
\jrg{The maximal speed is dictated by the thermodynamic stability of the final state controlled by the binding energy $\beta\epsilon_{b}$, Fig.~\ref{fig3}(a).
The minimal dissipation is largely dictated by the nonequilibrium currents around the intermediate states controlled by the kinetic coefficients $d/c$, Fig.~\ref{fig3}(b).}
To simultaneously minimize dissipation and maximize speed, we use the fact that the parameters provide independent control and extremize each observable in sequence.
Fig.~\ref{fig4}(a-bottom) shows that the minimum average entropy dissipated increases exponentially with $\beta\epsilon_{b}$, and with different growth rates on either side of $\beta\epsilon_{b}=-\ln c$.
We see in Fig.~\ref{fig3}(b) that this minimum entropy dissipated occurs for a given value of $\beta\epsilon_{b}$ at $c=d$.
At approximately the same relative binding energy, the associated average rate has a maximum, Fig.~\ref{fig4}(a-top).
This maximum suggests that the most optimal combination of extremizing both the entropy dissipated and the rate is obtained with a homogeneous kinetic network, $c=d=e^{-\beta\epsilon_{b}}$.
Similarly, the maximal average rate occurs for $\log d/c<0.4$, Fig.~\ref{fig4}(b-bottom).
We see from Fig.~\ref{fig3}(a) that the maximum average rate for a given $d$ occurs at $c=e^{-\beta\epsilon_{b}}$.
Within this range the associated entropy dissipated has a minimum at $c\approx d$, Fig.~\ref{fig4}(b-top).
Overall then, for this model, the minimization of the mean entropy dissipated and maximization of the mean speed at this level commute.

\textit{Conclusions.--}
Natural and synthetic systems often balance the speed of structure formation and the associated cost of dissipation.
Understanding this balance and the experimental factors causing an imbalance, particularly for transient behavior, has the potential for creating assembling systems that are adaptive and responsive to their environment.
\jrg{To address this problem here, we derived an analytical formula for the occurrence probability of stochastic paths through the path summation solution of the master equation.}
\jrg{This formula provides a means to identify and assess the kinetic relevance of paths for stochastic processes over a specified timeframe and without necessarily constraining endpoints.}
\jrg{This contracted path probability is necessary to directly calculate ensemble-level observables, such as entropy production and flow, from a set of paths with the explicit determination of the time to complete the process.}
Applying this formula to a model for self-assembly showed that increasing speed need not be accompanied by increasing dissipation.
Further, these results suggest that networks governing the dynamics of other physical and chemical systems, such as biochemical reaction cycles \jrg{and the dissipative cycles of chemically-active materials}, might be finely tuned to control whether faster paths dissipate less than slower paths on average, reversing the intuition that ``more haste'' brings ``more waste''.

\begin{acknowledgments}
\textit{Acknowledgements.--}
This material is based upon work supported by the National Science
Foundation under Grant No.~1856250 and the John Templeton Foundation.
We acknowledge the use of the supercomputing facilities managed by the Research
Computing Group at the University of Massachusetts Boston as well as the
University of Massachusetts Green High-Performance Computing Cluster.
The work of R.A.B.\ was supported in part by College of Science and Mathematics Dean's Doctoral Research Fellowship through fellowship support from Oracle, project ID R20000000025727.
\end{acknowledgments}


%

\end{document}


\title{Supplemental Material: Stochastic paths controlling speed and dissipation}

\author{Rebecca~A.~Bone}
\affiliation{Department of Chemistry,\
  University of Massachusetts Boston,\
  Boston, MA 02125, USA
}

\author{Daniel~J.~Sharpe}
\author{David~J.~Wales}
\affiliation{Department of Chemistry,\
University of Cambridge,\
Lensfield Road, Cambridge CB2 1EW, UK
}

\author{Jason~R.~Green}
\email[]{jason.green@umb.edu}
\affiliation{Department of Chemistry,\
  University of Massachusetts Boston,\
  Boston, MA 02125, USA
}
\affiliation{Department of Physics,\
  University of Massachusetts Boston,\
  Boston, MA 02125, USA
}

\maketitle




\subsection*{Contracted path probability}

In a continuous-time, discrete-state Markovian system with $N$ states and time-independent transition rates $w(x_{j}\vert x_{i})$,  the master equation, $\dot{\boldsymbol{p}}(t) = (\boldsymbol{Q}-\boldsymbol{D})\,\boldsymbol{p}(t)$~\cite{zwanzig2001nonequilibrium}, governs the evolution of the marginal probability distribution, $\boldsymbol{p}$.
Transition rates $(\boldsymbol{Q})_{y,x}=w(y \vert x)\geq 0$ from state $x$ to state $y$ populate the time-independent rate matrix, $\boldsymbol{Q}-\boldsymbol{D}$, with the escape rates along the diagonal of $(\boldsymbol{D})_{x,x}=w_{x_{i}}:=\sum_{j=1,i\neq j}^{N} w(x_{j} \vert x_{i})\geq 0$~\cite{Sun2006}.
The waiting time in each state is exponentially distributed,
\begin{equation}
\rho(\Delta t_{i} \vert x_{i}) = \rho_{i}=w_{x_{i}} e^{-\Delta t_{i} w_{x_{i}}},
\end{equation}
with respect to the escape rate $w_{x_{i}}$ from the state being occupied $x_{i}$ and independent of the other escape rates and states.  
The convolution of these exponential distributions along a path yields the probability of a path taking a certain amount of time $\tau$ to complete given that the path occurs,
\begin{equation}
p(\tau \vert \mathcal{C}_{n}) = \rho_{0}*\left[\rho_{1}*\left[\ldots*\left[\rho_{n-1}*\left[ e^{-w_{x_{n}}(t-t_{n})}\right]\right]\right]\right].
\label{convolutions}
\end{equation} 
The final function in these convolutions differs as it represents the survival probability in the final state from which there is no additional transition rather than the distribution of waiting times for which there is a transition.
We then make use of the probability that the path occurs,
\begin{equation}
p(\mathcal{C}_{n}) = p(x_{0}) \sum\limits_{i=1}\limits^{n} \frac{w(x_{i} \vert x_{i-1})}{w_{x_{i-1}}},
\end{equation}
to achieve the contracted path probability through Bayes' theorem $\mu(\mathcal{C}_{n},\tau)=p(\tau \vert \mathcal{C}_{n})p(\mathcal{C}_{n})$.

The non-explicit form of this contracted path probability is
\begin{align}
&\mu(\mathcal{C}_{n},\tau) = p(x_{0},t_{0})\prod\limits_{i=0}\limits^{n-1} w(x_{i+1} \vert x_{i}) \nonumber \\
&\times\sum\limits_{j=1}\limits^{n'} \nu_{j} \frac{\partial^{m_{j}-1}}{\partial w_{x_{j}}^{m_{j}-1}}  \left[ \frac{e^{-w_{x_{j}}\tau}}{\prod\limits_{k=1,\neq j}\limits^{n'} (w_{x_{k}}-w_{x_{j}})^{m_{k}}} \right].
\label{nonexplicit}
\end{align}
The prefactor (line 1 of Eqn.~\ref{nonexplicit}) is the result of the multiplication of $p(\mathcal{C}_{n})$ and the prefactor resulting from the use of Cauchy's Residue Theorem in deriving this non-explicit expression~\cite{Sun2006} or, equivalently, the prefactor for  $p(\tau \vert \mathcal{C}_{n})$, $\prod_{i=0}^{n} w_{x_{i}}$ that follows from evaluating the convolutions.
The following function $\nu_{j}$ corrects for the multiplicity of the $j^{\textrm{th}}$ unique escape rate in counting the number of identical derivatives correspond to the $j^{\textrm{th}}$ escape rate:
\begin{equation}
\nu_{j}=\frac{(-1)^{m_{j}-1}}{(m_{j}-1)!}
\end{equation}
The remaining sum yields one term for each state along the path whose escape rate is unique, of which there are $n'$.
These unique escape rates have an associated degeneracy $m_{j}$, the number of times that escape rate occurs along the path (e.g., an escape rate that occurs only once has a degeneracy of $m_{j}=1$). 
Escape rates along a path can be degenerate if more than one state along the path has the same escape rate (e.g., $w_{x_{i}}=w_{x_{j}}$), regardless of if those states are the same.

A fully degenerate path, a path in which all escape rates are the same, has only one term in this sum, whereas a non-degenerate path, a path in which no escape rates are the same, has $n'=n+1$ terms, one for each state along the path. 
These bounding cases have known simplifications of the contracted path probability formula because the form of the derivatives is known. 
The contracted path probability of a fully degenerate path has the form
\begin{equation}
\mu(\mathcal{C}_{n},\tau)=p(x_{0},t_{0}) \frac{\tau^{n}e^{-w\tau}}{n!} \prod\limits_{i=1}\limits^{n} w(x_{i} \vert x_{i-1}),
\end{equation}
which is similar to the Erlang distribution, and the form for a non-degenerate path is
\begin{equation}
\mu(\mathcal{C}_{n},\tau)=p(x_{0},t_{0}) \prod\limits_{i=1}\limits^{n} w(x_{i} \vert x_{i-1}) \sum\limits_{j=0}\limits^{n} \frac{e^{-w_{x_{j}}\tau}}{\prod\limits_{k=0,k\neq j}\limits^{n} (w_{x_{k}}-w_{x_{j}})}.
\end{equation}

\subsection*{Closed form expression of contracted path probability}

We have constructed an explicit general formulation of this contracted path probability:
\begin{equation}
  \mu(\mathcal{C}_{n},\tau ) = p(x_{0},t_{0}) \prod\limits_{i=0}\limits^{n-1} w(x_{i+1} \vert x_{i}) \sum\limits_{j=1}\limits^{n'} \left( \frac{(-1)^{m_{j}-1}}{(m_{j}-1)!} \right) \left( \frac{e^{-w_{x_{j}}\tau}}{\prod\limits_{\substack{k=1 \\ \neq j}}\limits^{n'} (w_{x_{k}}-w_{x_{j}})^{m_{k}}} \right) \sum\limits_{l=1}\limits^{m_{j}} \binom{m_{j}-1}{l-1} (-\tau )^{m_{j}-l} d_{j}^{(l-1)}. \label{probability}
\end{equation}
To derive this result, we first recognize that the only portion of the non-explicit formula whose explicit counterpart is unknown is the form of the successive derivatives of the function
\begin{equation}
f_{j}^{(0)}=\frac{e^{-w_{x_{j}}\tau}}{\prod\limits_{k=1,\neq j}\limits^{n'} (w_{x_{k}}-w_{x_{j}})^{m_{k}}}
\end{equation}
with respect to the $j^{\textrm{th}}$ unique escape rate $w_{x_{j}}$. 
These derivatives can be stated as a product rule $f_{j}^{(0)}=g(w_{x_{j}})h(w_{x_{j}})$ where our two functions are the exponential in the numerator $g(w_{x_{j}})=e^{-w_{x_{j}}\tau}$ and the product in the denominator $h(w_{x_{j}})=\left[ \prod_{k=1,k\neq j}^{n'} (w_{x_{k}}-w_{x_{j}})^{m_{k}} \right]^{-1}$. 
Due to the product in $h(w_{x_{j}})$, each derivative of $f_{j}$ results in a series of terms, all of which are multiplied by the initial function $f_{j}^{(0)}$. 
Each successive derivative of this function accumulates a series of terms $\tau^{0}$ to $\tau^{m_{j}-1}$ of alternating sign for the $l^{\textrm{th}}$ derivative of $g(w_{x_{j}})$. 
The terms of order $\tau^{0}$ and $\tau^{m_{j}-1}$ correspond to the pure derivatives of the denominator and the numerator respectively. 
The $q^{\textrm{th}}$ pure derivative of the numerator has the form $(-\tau)^{q}e^{-w_{x_{j}}\tau}$. 

%
%

The pure derivative of the denominator has a form that is harder to access. 
Each index of the product within $h(w_{x_{j}})$ has its own chain rule, and the combined product of which $h(w_{x_{j}})$ consists has a product rule relating each of the terms of the product.
The form of the $l^{\textrm{th}}$ derivative resulting from the chain rule for the $k^{\textrm{th}}$ index of the product is
\begin{equation}
\frac{\partial^{l}}{\partial w_{x_{j}}^{l}} \left( \frac{1}{(w_{x_{k}}-w_{x_{j}})^{m_{k}}}\right) =\frac{(-1)^{l}\prod\limits_{u=0}\limits^{l-1} (m_{k}+u)}{(w_{x_{k}}-w_{x_{j}})^{m_{k}+l}}.
\end{equation}
For the first derivative of $h(w_{x_{j}})$, this is sufficient to determine the generalized form
\begin{equation}
h^{(1)}(w_{x_{j}})= \left( \frac{1}{\prod\limits_{k=1,\neq j} (w_{x_{k}}-w_{x_{j}})^{m_{k}}} \right) (-1) \sum\limits_{l=1,\neq j}\limits^{n'} \frac{m_{l}}{w_{x_{l}}-w_{x_{j}}}.
\end{equation}
The sum accounts for the derivative of each term in the product.

With each additional derivative taken of $h(w_{x_{j}})$, we must then take into account all combinations of orders of derivatives among the terms in the product. 
In each combination, the order of the derivatives of each term in the product must sum to the total order of the derivative of $h(w_{x_{j}})$. 
The original function $h(w_{x_{j}})$ is an eigenfunction with respect to the derivative operator, just as every derivative of $g(w_{x_{j}})$ contains $g(w_{x_{j}})$: $\partial^{k} / \partial w_{x_{j}}^{k} \left( e^{-w_{x_{j}}\tau} \right) = (-\tau)^{k} e^{-w_{x_{j}}\tau}$. 
Thus, each derivative of $f_{j}$ contains the original function $f_{j}^{(0)}$.
The eigenvalue then accounts for the multiplicities of the unique escape rates and the mixing terms for combinations of these unique escape rates.
The remaining $m_{j}-3$ terms of index $l$ account for mixed derivatives of both the numerator and denominator. 
The number of occurrences of each of these mixing terms is represented by the binomial coefficient $\binom{m_{j}-1}{l-1}$.
Each mixing term then has the product of the $(l-1)^{\textrm{th}}$ derivative of $g(w_{x_{j}})$ and the $(m_{j}-l)^{\textrm{th}}$ derivative of $h(w_{x_{j}})$.

The fully explicit equation resulting from this is shown in Eqn.~\ref{probability}. 
The function $d_{j}^{(l-1)}$ holds the form of the $h(w_{x_{j}})$ portion of the eigenvalue of $f_{j}$ (i.e., the portion of $f_{j}^{(m_{j}-1)} / f_{j}^{(0)}$ corresponding to the derivative(s) of the denominator).
The structure of the $k^{\textrm{th}}$ term in a single function $d_{j}^{(l-1)}$ can be thought of as one of the unrestricted partitions of $l-1$.
The number of terms in $d_{j}^{(l-1)}$  is also the number of unrestricted partitions of $l-1$: $\vert U_{l-1} \vert$. 
For the $q^{\textrm{th}}$ partition, the number of coefficients $\chi$ in the partition $\vert U_{k-1}^{q} \vert$ is the number of sums for that term. 
The sum is evaluated over each possible combination of unique escape rates $w_{x_{v}}\neq w_{x_{j}}$ along the path.
Each sum term corresponding to one combination of escape rates is then comprised of a product of fractions.
Each fraction has a numerator portion containing the terms with the multiplicity of that escape rate and a denominator with a difference term between the escape rate and the $j^{\textrm{th}}$ escape rate to the power of the number of terms of that escape rate's multiplicity in the numerator.
The $k^{\textrm{th}}$ term in a function $d_{j}^{(l-1)}$ is only evaluated when there are enough unique escape rates other than $j$ to have an escape rate for each sum forming that term. 
Thus, a path with only three unique escape rates would not evaluate the final term of $d_{j}^{(3)}$ as there are only two unique escape rates other than $j$. 

The coefficient of the $k^{\textrm{th}}$ term in $d_{j}^{(l-1)}$ is also related to the unrestricted partitions of $l-1$ by the formula
\begin{equation}
C(l-1,k)=\frac{(l-1)!}{\prod\limits_{v=1}\limits^{\vert U_{l-1}^{k} \vert} (\chi_{v})! \prod\limits_{w=1}\limits^{\vert U_{l-1}^{k} \vert '} (\lambda_{w})!}
\end{equation}
For this formula, we further need to know that there are $\vert U_{l-1}^{q} \vert'$ unique values in the $q^{\textrm{th}}$ partition of $l-1$.
The $v^{\textrm{th}}$ unique value in the $q^{\textrm{th}}$ partition of $l-1$ has a multiplicity of $\lambda_{v}$.
From this, we construct the first few functions $d_{j}^{(l-1)}$. For $l-1=1$, the form of this function is
\begin{equation}
d_{j}^{(1)}= \sum\limits_{\alpha=1,\neq j}\limits^{n'} \frac{m_{\alpha}}{w_{x_{\alpha}}-w_{x_{j}}}
\end{equation}
This is expected from the form of the first derivative of $h(w_{x_{j}})$ detailed above. 

For $l-1=2$, the form of this function becomes
\begin{align}
d_{j}^{(2)}&= \sum\limits_{\alpha=1,\neq j}\limits^{n'} \frac{m_{\alpha}(m_{\alpha+1})}{(w_{x_{\alpha}}-w_{x_{j}})^{2}} \nonumber \\
&+ \sum\limits_{\beta=1,\neq j}\limits^{n'} \sum\limits_{\gamma=1,\neq j,\beta}^{n'} \frac{m_{\beta}m_{\gamma}}{(w_{x_{\beta}}-w_{x_{j}})(w_{x_{\gamma}}-w_{x_{j}})}
\end{align}
The first term corresponds to the partition (2) and has coefficient $C(2,2)=1$. 
The second term corresponds to the partition (1,1) and has coefficient $C(2,1)=1$. 

Additional examples of $d_{j}^{(l-1)}$ are as follows. 
For $l-1=3$, the function is given by 
\begin{align}
d_{j}^{(3)} &= \sum\limits_{\substack{\alpha = 1 \\ \neq j}}\limits^{n'} \frac{m_{\alpha}(m_{\alpha}+1)(m_{\alpha}+2)}{(w_{x_{\alpha}}-w_{x_{j}})^{3}} + 3\sum\limits_{\substack{\beta = 1 \\ \neq j}}\limits^{n'} \sum\limits_{\substack{\gamma = 1 \\ \neq j,\beta}}\limits^{n'} \frac{m_{\beta}(m_{\beta}+1)m_{\gamma}}{(w_{x_{\beta}}-w_{x_{j}})^{2}(w_{x_{\gamma}}-w_{x_{j}})} + \sum\limits_{\substack{\delta = 1 \\ \neq j}}\limits^{n'} \sum\limits_{\substack{\epsilon = 1 \\ \neq j,\delta}}\limits^{n'} \sum\limits_{\substack{\zeta = 1 \\ \neq j,\delta ,\epsilon}}\limits^{n'} \frac{m_{\delta}m_{\epsilon}m_{\zeta}}{(w_{x_{\delta}}-w_{x_{j}})(w_{x_{\epsilon}}-w_{x_{j}})(w_{x_{\zeta}}-w_{x_{j}})}.
\end{align}
This function corresponds to the third row of Fig.~3(c). 
Term 1 corresponds to the partition (3) with coefficient $C(3,3)=1$. 
Term 2 corresponds to the partition (2,1) with coefficient $C(3,2) = 3$. 
Term 3 corresponds to the partition (1,1,1) with coefficient $C(3,1)=1$.

For $l-1=4$, the function is given by
\begin{align}
d_{j}^{(4)} &= \sum\limits_{\substack{\alpha = 1 \\ \neq j}}\limits^{n'} \frac{m_{\alpha}(m_{\alpha}+1)(m_{\alpha}+2)(m_{\alpha}+3)}{(w_{x_{\alpha}}-w_{x_{j}})^{4}} + 4 \sum\limits_{\substack{\beta = 1 \\ \neq j}}\limits^{n'} \sum\limits_{\substack{\gamma = 1 \\ \neq j,\beta}}\limits^{n'} \frac{m_{\beta}(m_{\beta}+1)(m_{\beta}+2)m_{\gamma}}{(w_{x_{\beta}}-w_{x_{j}})^{3}(w_{x_{\gamma}}-w_{x_{j}})} + 3 \sum\limits_{\substack{\delta = 1 \\ \neq j}}\limits^{n'} \sum\limits_{\substack{\epsilon = 1 \\ \neq j,\delta}}\limits^{n'} \frac{m_{\delta}(m_{\delta}+1)m_{\epsilon}(m_{\epsilon}+1)}{(w_{x_{\delta}}-w_{x_{j}})^{2}(w_{x_{\epsilon}}-w_{x_{j}})^{2}} \nonumber \\
&+ 6 \sum\limits_{\substack{\zeta = 1 \\ \neq j}}\limits^{n'} \sum\limits_{\substack{\eta = 1 \\ \neq j,\zeta}}\limits^{n'} \sum\limits_{\substack{\theta = 1 \\ \neq j,\zeta,\eta}}\limits^{n'} \frac{m_{\zeta}(m_{\zeta}+1)m_{\eta}m_{\theta}}{(w_{x_{\zeta}}-w_{x_{j}})^{2}(w_{x_{\eta}}-w_{x_{j}})(w_{x_{\theta}}-w_{x_{j}})} \nonumber \\
&+ \sum\limits_{\substack{\iota = 1 \\ \neq j}}\limits^{n'} \sum\limits_{\substack{\kappa = 1 \\ \neq j,\iota}}\limits^{n'} \sum\limits_{\substack{\lambda = 1 \\ \neq j,\iota,\kappa}}\limits^{n'} \sum\limits_{\substack{\mu = 1 \\ \neq j,\iota,\kappa,\lambda}}\limits^{n'} \frac{m_{\iota}m_{\kappa}m_{\lambda}m_{\mu}}{(w_{x_{\iota}}-w_{x_{j}})(w_{x_{\kappa}}-w_{x_{j}})(w_{x_{\lambda}}-w_{x_{j}})(w_{x_{\mu}}-w_{x_{j}})}.
\end{align}
This function corresponds to the fourth row of Fig.~3(c). 
Term 1 corresponds to the partition (4) and has coefficient $C(4,5)=1$. 
Term 2 corresponds to the partition (3,1) and has coefficient $C(4,4)=4$. 
Term 3 corresponds to the partition (2,2) and has coefficient $C(4,3)=3$. 
Term 4 corresponds to the partition (2,1,1) and has coefficient $C(4,2)=6$. 
Term 5 corresponds to the partition (1,1,1,1) and has coefficient $C(4,1)=1$.

\subsection*{Validation}

The form of the explicit contracted path probability formula was confirmed by comparison of symbolic and calculated evaluations of the convolution integral form obtained by multiplying Eqn.~\ref{convolutions} by the path probability $p(\mathcal{C}_{n})$, the non-explicit contracted path probability formula in~\cite{Sun2006}, and the explicit contracted path probability. 
This by-hand confirmation was carried out for all possible degeneracies of paths of lengths $n\leq10$.
Further, collapse to the fully degenerate and non-degenerate forms of the explicit contracted path probability formula were confirmed analytically.

To further validate the resulting contracted path probability formula, we generated a sample of stochastic trajectories for a path of length $n=1000$~\cite{Athenes2014}. 
The histogram of the occurrence time of these trajectories agreed with the analytical distribution of occurrence times for the trajectories following a path.
This analytical distribution is given by
\begin{align}
\rho(\tau \vert \mathcal{C}_{n}) &= \prod\limits_{i=0}\limits^{n-1} w_{x_{i}} \sum\limits_{j=1}\limits^{n'} \left( \frac{(-1)^{m_{j}-1}}{(m_{j}-1)!} \right) \left( \frac{e^{-w_{x_{j}}\tau}}{\prod\limits_{\substack{k=1 \\ \neq j}}\limits^{n'} (w_{x_{k}}-w_{x_{j}})^{m_{k}}} \right) \nonumber \\
&\times \left(\sum\limits_{l=1}\limits^{m_{j}} {m_{j}-1 \choose l-1} (-\tau )^{m_{j}-l} d_{j}^{(l-1)} \right).
\label{dist_occ_time}
\end{align}
The mean of this distribution is given by
\begin{equation}
\langle \tau \rangle_{\mathcal{C}_{n}} = \int_{t_{0}}^{\infty} d\tau \tau \rho(\tau \vert \mathcal{C}_{n})=\sum_{i=0}^{n} \frac{1}{w_{x_{i}}}.
\end{equation}
This can alternatively be derived by assumption that the exponentially-distributed waiting times along a path are independent. 
The mean is then recovered by summing the means for each exponential distribution of waiting times. 
The mean path occurrence time calculated from the analytical distribution is situated at the peak of both the empirical and theoretical distributions. 

\subsection*{Models}

There are regions of the parameter space where thermodynamic intuition does not hold -- regions in which paths that complete more quickly dissipate less, Fig.~2(a) of main text.
This physical scenario occurs when the dissipation of the cycle is significant, $c\neq d$.
Under these conditions, longer paths have sufficient length to undergo more transitions around the cycle and, so, dissipate more.
Shorter paths, however, have less opportunity to transition around the cycle as they visit fewer states and, so, take less time to reach the target state.
While shorter paths cannot exchange as much entropy with the environment and complete quickly, longer paths exchange more entropy with the environment and take longer to complete.
This correlation ensures that shorter paths are faster and dissipate less and that longer paths are slower and dissipate more: speed and dissipation are inversely related.

\begin{figure}[ht]
\centering
\includegraphics[width=1\textwidth]{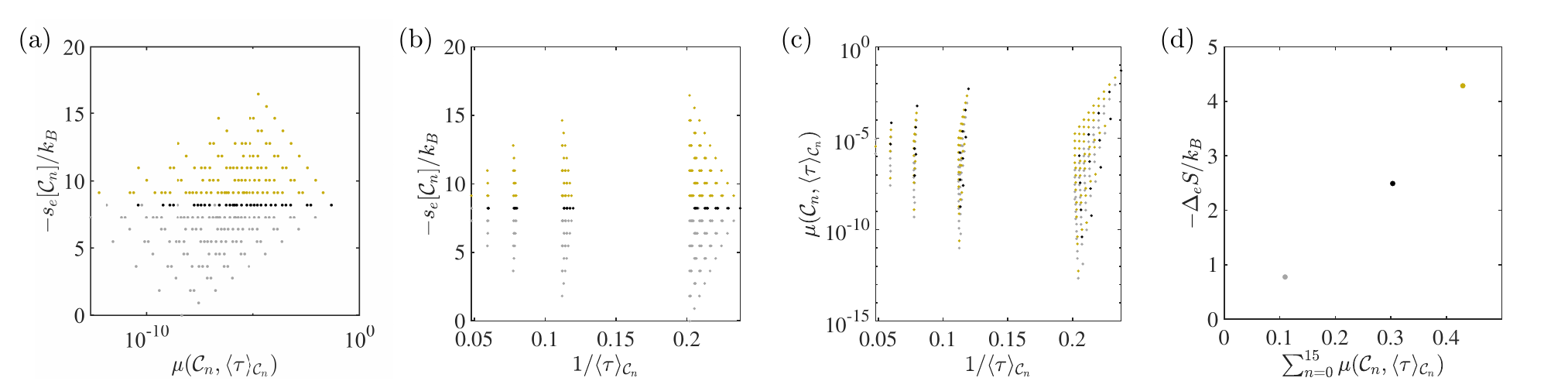}
\caption{ \label{SI1} 
(a) Path-level entropy flow versus the contracted path probability with paths net following the favored cycle direction in gold, paths net following the unfavored cycle direction in gray, and paths with no net cycle traversing in black. 
(b) Path-level entropy flow versus rate of path completion with paths net following the favored cycle direction in gold, paths net following the unfavored cycle direction in gray, and paths with no net cycle traversing in black. 
(c) Path-level contracted path probability versus rate of path completion with paths net following the favored cycle direction in gold, paths net following the unfavored cycle direction in gray, and paths with no net cycle traversing in black.
(d) Average over paths net following the favored cycle direction in gold, paths net following the unfavored cycle direction in gray, and paths with no net cycle traversing in black of negative of the entropy flow (left axis) vs. sum of contracted path probability and of the rate of path completion (right axis) vs. sum of contracted path probability.
Parameters for all plots are: $c=5$, $d=2$, $\beta\epsilon_{b}=5$.}
\end{figure}

\begin{figure}[ht]
\centering
\includegraphics[width=0.8\textwidth]{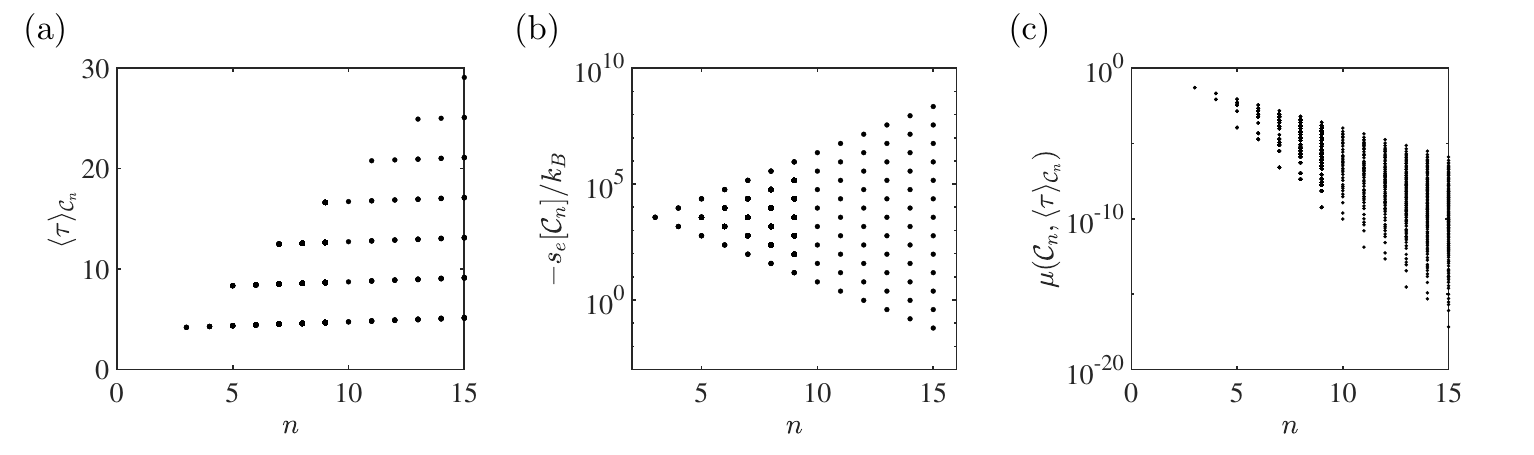}
\caption{ \label{SI2}
(a) Path occurrence time vs. path length $n$. Red, dashed lines indicate linear bounds with growing $n$.
(b) Path-level entropy flow vs. path length $n$. Red, dashed lines indicate linear bounds with growing $n$.
(c) Contracted path probability vs. path length $n$. Red, dashed lines indicate exponential bounds with growing $n$.
Parameters for all plots are: $c=5$, $d=2$, $\beta\epsilon_{b}=5$.}
\end{figure} 

\begin{figure}[ht]
\centering
\includegraphics[width=1\textwidth]{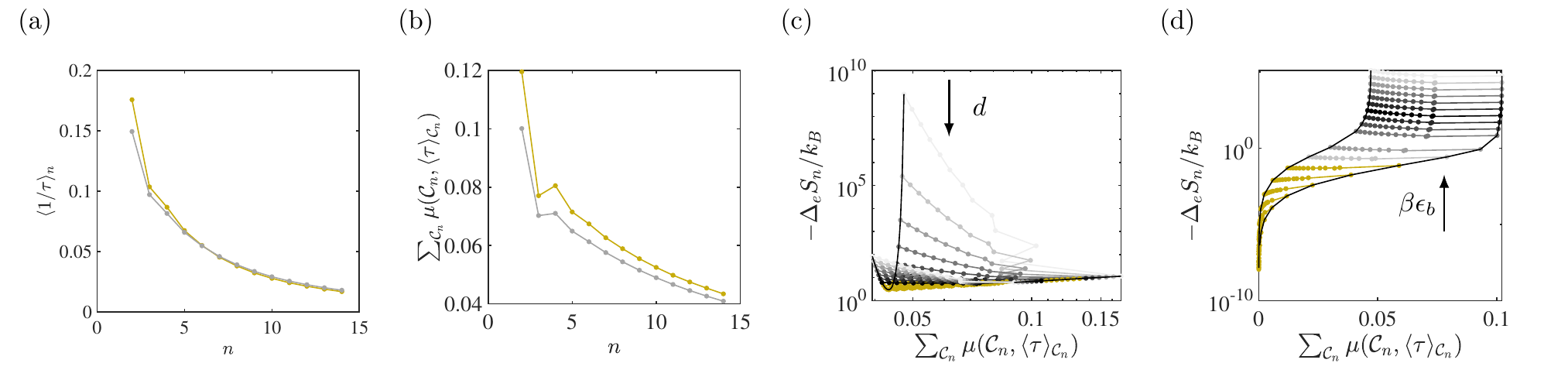}
\caption{ \label{SI3}
(a) Path-ensemble level mean rate of path completion vs. path length $n$. Gold line indicates trend with no cycle dissipation $c=d=5$. Gray line indicates trend with cycle dissipation $c=5$, $d=8$.
(b) Path-ensemble level yield (sum of contracted path probabilities) vs. path length $n$. Gold line indicates trend with no cycle dissipation $c=d=5$. Gray line indicates trend with cycle dissipation $c=5$, $d=8$. Both have $\beta\epsilon_{b}=1$.
(c) The dissipated entropy, $-\Delta_{e} S_{n}$, shows similar trends as a function of the total contracted path probability $p(\mathcal{C}|n):=\sum_{\mathcal{C}_n} p\left(\mathcal{C}_{n},\langle\tau\rangle_{\mathcal{C}_{n}}\right)$ for fixed-endpoint paths of length $n$.
What controls the slope is the current around the cycle, which we tune by sweeping $d\in [0.5,10]$ in increments of 0.5 at fixed $c=5$ and $\beta\epsilon_{b}=1$.
Dissipation increases with the rate of path completion (gold) for $c\approx d$ but decreases sharply when $|c-d|\gg 0$ (grey).
Lines connecting the slowest paths for the parameter sweep shows a parabolic trend (black).
(d) Dissipation, $-\Delta_{e} S_{n}$, as a function of the total contracted path probability $p(\mathcal{C}|n):=\sum_{\mathcal{C}_n} p\left(\mathcal{C}_{n},\langle\tau\rangle_{\mathcal{C}_{n}}\right)$ resolved by the path length $n$.
The slope is controlled by sweeping $\beta\epsilon_{b}\in[-10,10]$ in increments of 1 with fixed $c=5$ and $d=8$.
Color scheme is the same as in (c).
}
\end{figure}

\clearpage
\begin{figure}[ht]
\centering
\includegraphics[width=1\textwidth]{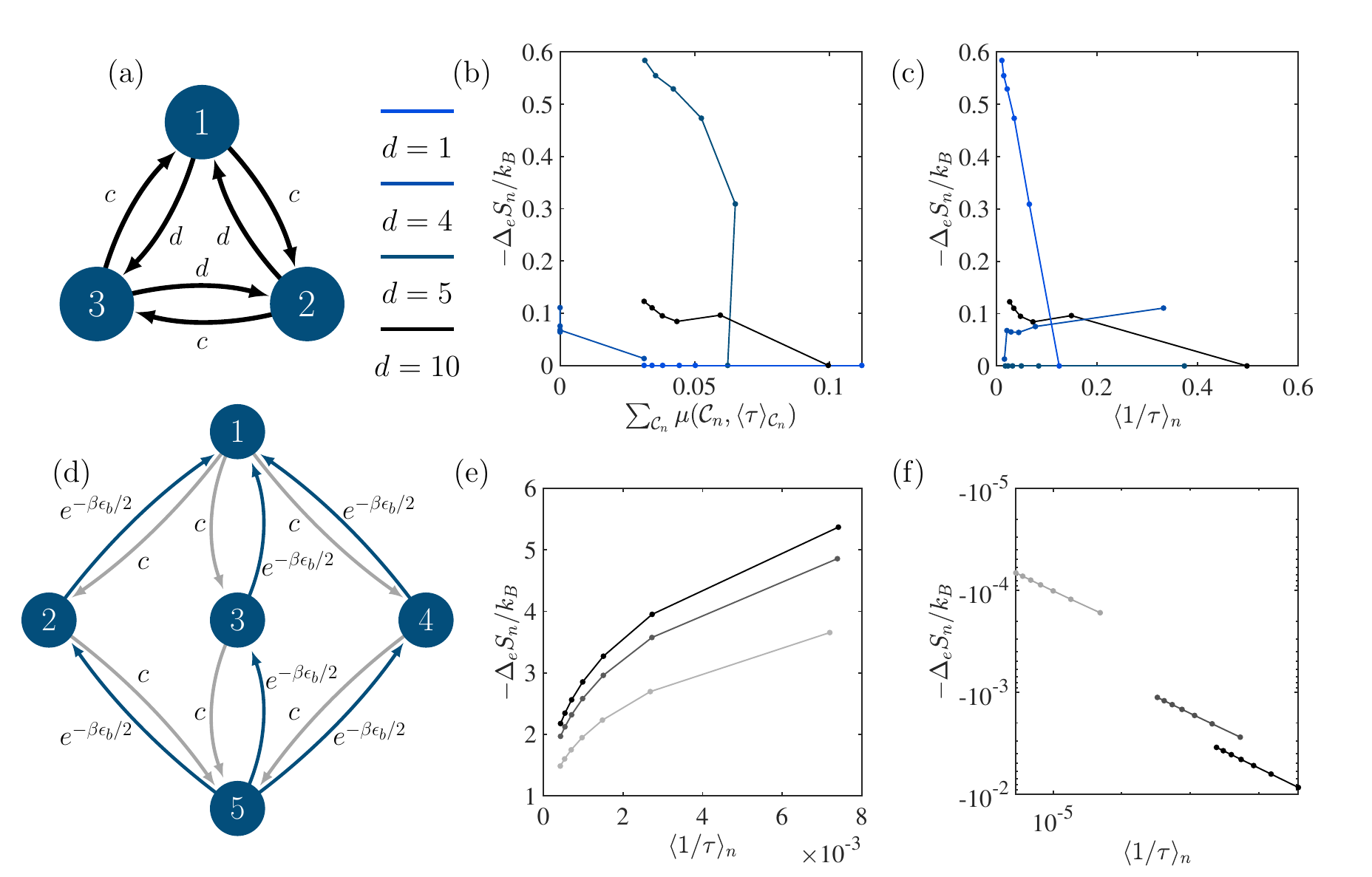}
\caption{ \label{entflow_vs_n}
(a) Diagram of model consisting of only the dissipating cycle.
(b) Entropy dissipated averaged over paths of length $n$ vs. yield of paths of length $n$ for the model in (a) with $c=5$ and $d=\lbrace 1, 4, 5, 10\rbrace$ (darker blue indicates higher $d$).
(c)  Entropy dissipated averaged over paths of length $n$ vs. mean rate of path completion of paths of length $n$ for the model in (a) with $c=5$ and $d=\lbrace 1, 4, 5, 10 \rbrace$ (darker blue indicates higher $d$).
(d) Diagram of model consisting of original model without cycle transitions.
(e) Entropy dissipated averaged over paths of length $n$ vs. mean rate of path completion for the model in (c) with $c=5$ and $\beta\epsilon_{b}=10$.
(f) Entropy dissipated averaged over paths of length $n$ vs. mean rate of path completion for the model in (c) with $c=5$ and $\beta\epsilon_{b}=-10$.}
\end{figure}


%